\documentclass[conference, anonymous]{IEEEtran}

\usepackage{amsmath,amssymb,amsfonts}
\usepackage{graphicx}
\usepackage{svg}
\usepackage{textcomp}
\usepackage{xcolor}
\usepackage{url}
\usepackage{booktabs,tabularx}
\usepackage{multirow}
\usepackage[caption=false,font=footnotesize]{subfig}
\usepackage{array}
\usepackage{longtable}

\usepackage{pifont}  
\usepackage{balance}
\usepackage{textcomp}
\usepackage{comment} 
\usepackage{float}
\usepackage{booktabs}
\usepackage{xcolor}
\usepackage{algorithm}
\usepackage{tabularx}
\usepackage{algpseudocode}
\usepackage{rotating}
\usepackage{wasysym}
\usepackage{subcaption}
\usepackage{afterpage}
\usepackage{enumitem}

\usepackage{stfloats}
\usepackage{comment} 
\usepackage{rotating}
\usepackage{placeins}

\AtBeginDocument{%
  \providecommand\BibTeX{{%
    Bib\TeX}}}
\def\BibTeX{{\rm B\kern-.05em{\sc i\kern-.025em b}\kern-.08em
    T\kern-.1667em\lower.7ex\hbox{E}\kern-.125emX}}

\begin{document}

%\title{Towards a Scalable AI-Driven Service for Analyzing User Engagement and Stance Analysis on Social Media Content}
\title{Scalable AI-Driven Analytics for User Engagement and Stance Detection on Social Media}

\author{
\IEEEauthorblockN{Thammitage Piyumi Wathsala Seneviratne, Muhammad Ikram, Dinusha Vatsalan,\\ Hassan Asghar, Mohamed Ali Kaafar}
\IEEEauthorblockA{
School of Computing, Macquarie University\\
\{thammitagepiyumiwathsala, muhammad.ikram, dinusha.vatsalan, hassan.asghar, dali.kaafar\}@mq.edu.au
}
}

\maketitle

\begin{abstract}
Social media platforms have become a major vector for the large-scale dissemination of misinformation and conspiracy content, posing significant risks to public trust, health, and societal stability. While prior work has primarily focused on analysing such content from a behavioural or content-centric perspective, there is a lack of scalable, service-oriented solutions that enable continuous monitoring and analysis of user engagement at platform scale.

In this paper, we present a scalable AI-driven service framework for analysing user engagement and stance on social media content. Our system integrates data ingestion, filtering, topic modelling, sentiment analysis, and stance detection into a modular pipeline that can operate on large-scale, real-world datasets. We implement and evaluate our framework on a dataset comprising over 7 million user comments collected from nearly 50,000 YouTube videos associated with conspiracy narratives.

Our analysis reveals that conspiracy content attracts up to 70\% of total user engagement within the first week of publication, indicating strong early amplification dynamics. Furthermore, we identify a subset of highly active users who exhibit disproportionately high engagement across multiple videos and channels. Stance analysis shows that a majority of users express favourable positions toward conspiracy narratives, highlighting the role of user communities in reinforcing such content.

The proposed framework demonstrates the feasibility of deploying scalable, service-oriented analytics for real-time monitoring of user engagement and behavioural patterns. These findings demonstrate the effectiveness of our framework in capturing large-scale engagement dynamics and highlight the importance of early-stage detection and service-based monitoring for mitigating the spread of harmful content.

\end{abstract}

\maketitle

\section{Introduction}
\label{sec:intro}
YouTube is one of the most widely used web platforms, with over 2.5 billion monthly users and more than 500 hours of content uploaded every minute~\cite{Top_Websites_Ranking_2023}\cite{blogYouTubestats}. While such platforms enable large-scale content sharing and interaction, they have also become major vectors for the dissemination of misinformation and conspiracy narratives. These content types have been shown to negatively influence public opinion, erode trust, and contribute to real-world societal harms, including resistance to public health measures and political instability~\cite{ROMER2020113356}\cite{dastgeer2022qanon}.

Despite ongoing improvements in content moderation and recommendation systems~\cite{blogManagingHarmful}\cite{improveRecommendations}, recent policy changes indicate a shift toward more permissive handling of certain forms of controversial content. This increases the need for scalable, automated mechanisms that can continuously monitor user engagement and detect emerging patterns associated with harmful or misleading content. Prior research has largely focused on understanding misinformation and conspiracy content from a behavioural or content-centric perspective~\cite{allcott2019trends}\cite{Faddoul2020}\cite{Boulianne2022}. However, there remains a critical gap in designing service-oriented, deployable solutions that enable platform-scale analysis of user engagement, sentiment, and stance.

In this paper, we address this gap by designing a scalable AI-driven web service framework for analysing user engagement and stance on social media content. Our system adopts a modular pipeline architecture that integrates data ingestion, large-scale comment processing, filtering, topic modelling, sentiment analysis, and stance detection into a unified service. The framework is designed to operate on real-world, high-volume data streams and can support continuous monitoring and analytics through service interfaces.

To demonstrate the effectiveness and scalability of our approach, we implement and evaluate the proposed framework on a large-scale dataset comprising over 7 million publicly available user comments collected from nearly 50,000 YouTube videos associated with conspiracy narratives~\cite{Ballard2022}. Our analysis reveals that conspiracy content attracts up to 70\% of user engagement within the first week of publication, indicating strong early amplification dynamics. Furthermore, we identify a subset of highly active users who exhibit disproportionately high engagement across multiple videos and channels, suggesting coordinated or persistent engagement behaviours. Our stance analysis further shows that a majority of users express favourable positions toward conspiracy narratives, highlighting the role of user communities in reinforcing such content.

The proposed framework provides a foundation for building next-generation web services that enable real-time monitoring, risk assessment, and content moderation support. By combining scalable system design with advanced NLP techniques, our work bridges the gap between empirical analysis and deployable service architectures for social media platforms. Our proposed framework aims to answer the following research questions:

\begin{itemize}[topsep=0pt]
\item {\bf $\mathbf{RQ1}$ (Engagement Analysis):} How do users engage with conspiracy content compared to mainstream content in terms of interaction patterns such as comments and likes?
\item {\bf $\mathbf{RQ2}$ (Behavioural Signals):} Which sentiments, emotions, and stances influence active user engagement?
\item {\bf $\mathbf{RQ3}$ (Temporal Dynamics):} How does user engagement evolve over time following content publication?
\end{itemize}

Our work makes the following key contributions:
\begin{enumerate}[topsep=0pt]
    \item We design and implement a scalable AI-driven web service framework for large-scale analysis of user engagement and stance on social media platforms.
    
    \item We develop a modular processing pipeline integrating filtering, topic modelling, sentiment analysis, and a weakly supervised stance detection approach tailored for large-scale, unlabelled social media data.
    
    \item We conduct a comprehensive empirical evaluation on a dataset of over 7 million comments, revealing key engagement patterns such as rapid early amplification and the presence of highly active user subsets.
    
    \item We provide insights into sentiment, emotion, and stance dynamics, showing that engagement with conspiracy content is strongly associated with negative sentiment and favourable stance toward such narratives.
    
    \item We demonstrate the feasibility of deploying service-oriented analytics for real-time monitoring and decision support in content moderation and platform governance.
\end{enumerate}

\section{Related Work}
\label{sec.Related Work}
In this section, we review prior work in two closely related areas: conspiracy content on social media broadly, and conspiracy content on YouTube specifically. We then position the gap addressed by our work from the perspective of scalable, service-oriented analysis.

\noindent
{\bf Conspiracy content on social media:}
Conspiracy narratives and misinformation have been widely studied across major social platforms. Prior research has shown that conspiracy consumers tend to be more concentrated in their interests and more likely to promote dubious information. For example, Bessi et al.~\cite{bessi2015science} found that Facebook users engaging with conspiracy content exhibit stronger concentration and reinforcement behaviours than consumers of science content. Del Vicario et al.~\cite{del2016spreading} further showed that conspiracy communities are characterised by strong polarisation and echo-chamber effects. Extending this line of work, Bessi et al.~\cite{Bessi2016} demonstrated that commenting behaviour itself contributes to the formation and reinforcement of such echo chambers across Facebook and YouTube.

Beyond Facebook, related work on Reddit has shown that conspiracy content spreads more virally than science content and is often associated with more negative emotional language~\cite{zhang2021conspiracy}. During the COVID-19 pandemic, conspiracy narratives intensified and became intertwined with public-health misinformation~\cite{shahsavari2020conspiracy}. More broadly, Cinelli et al.~\cite{cinelli2022conspiracy} argued that conspiracy ecosystems across platforms are shaped by polarisation, moderation asymmetries, and migration toward less regulated spaces.

Taken together, these studies establish that conspiracy content is socially reinforced, emotionally charged, and structurally amplified across platforms. However, most of this literature focuses primarily on diffusion patterns, community structure, or content characteristics, rather than on building scalable analytical pipelines that jointly capture engagement behaviour, behavioural signals, and stance at platform scale.

\noindent
{\bf Conspiracy content on YouTube:}
YouTube has received particular attention because of its scale, recommendation mechanisms, and the persistence of conspiratorial material on the platform. Existing research has examined both content characteristics and user interaction patterns, but much of it remains limited in scope. Hussain et al.~\cite{hussain2018analyzing}, for instance, analysed a single YouTube channel focused on World War III conspiracies and found spikes in engagement and bot-like activity, but did not investigate the sentiments or emotions expressed in user comments. Röchert et al.~\cite{rochert2022caught} studied communication around three specific conspiracy theories---``Hollow Earth'', ``Chemtrails'', and ``New World Order''---and showed that pro-conspiracy videos attracted stronger engagement from supportive users. Their work, however, was constrained to a small set of theories and did not differentiate broader engagement dynamics at scale.

Faddoul et al.~\cite{Faddoul2020} provided an important longitudinal view of YouTube’s recommendation ecosystem by analysing 8 million recommended videos and showing that platform interventions reduced recommendation of conspiracy videos. Nevertheless, their comment analysis was limited to the top 200 comments per video and excluded reply structures, which constrains understanding of user interaction and conversational stance. Allington et al.~\cite{Allington2021} examined antisemitic conspiracy content and showed that harmful narratives persisted in comments even when videos were removed, but their study was centred on a narrow thematic subset. Ha et al.~\cite{Ha2022} took a complementary perspective by examining conspiratorial comments in mainstream news videos, especially those mentioning Bill Gates, and found that shorter conspiracy-oriented comments attracted high engagement. Ballard et al.~\cite{Ballard2022} shifted attention to monetisation, showing that predatory advertisements appeared substantially more often on conspiracy videos than on mainstream videos.

Although these studies provide valuable insights, they exhibit three key limitations. First, many focus on one or a few specific conspiracy theories, channels, or topical niches, which limits generalisability. Second, several studies analyse only a subset of available comments, often excluding reply threads, thereby overlooking richer user interaction structures. Third, prior work is predominantly analytical rather than service-oriented: it characterises content and communities, but does not provide a unified, scalable framework for continuous engagement analysis, behavioural signal extraction, and stance inference.

\noindent
{\bf Gap and Positioning of Our Work:}
Our work addresses this gap by moving beyond topic-specific or channel-specific analyses toward a scalable, AI-driven framework for engagement analysis on social media content. Unlike prior YouTube studies that focus on a single theory, a small set of channels, or a limited comment sample, we analyse all publicly available comments and replies associated with a large corpus of conspiracy videos spanning hundreds of channels and approximately 180 conspiracy topics. This broader coverage enables us to characterise engagement patterns at scale, rather than within isolated communities.

More importantly, our work differs from existing literature in both scope and system design. At the methodological level, we combine engagement analysis, sentiment and emotion modelling, topic-aware stance detection, and temporal analysis within a unified processing pipeline. At the systems level, we frame these capabilities as components of a scalable, service-oriented architecture that can support large-scale monitoring, early detection, and moderation-oriented analytics. In this sense, our contribution is not only empirical but also infrastructural: we provide a deployable analytical framework that bridges social media measurement and web-service-based behavioural intelligence.

%%%%%%%%% ################ Methodology Section Here
\section{System Design and Evaluation}
\label{sec:method}
In this section, we present the design, implementation, and evaluation of our scalable AI-driven web service framework for analysing user engagement and stance on social media content. Unlike traditional analysis pipelines, our approach is structured as a modular, service-oriented architecture that supports scalable processing and near real-time analytics.

\begin{figure*}[th]
\centering
\includegraphics[width=0.9\textwidth]{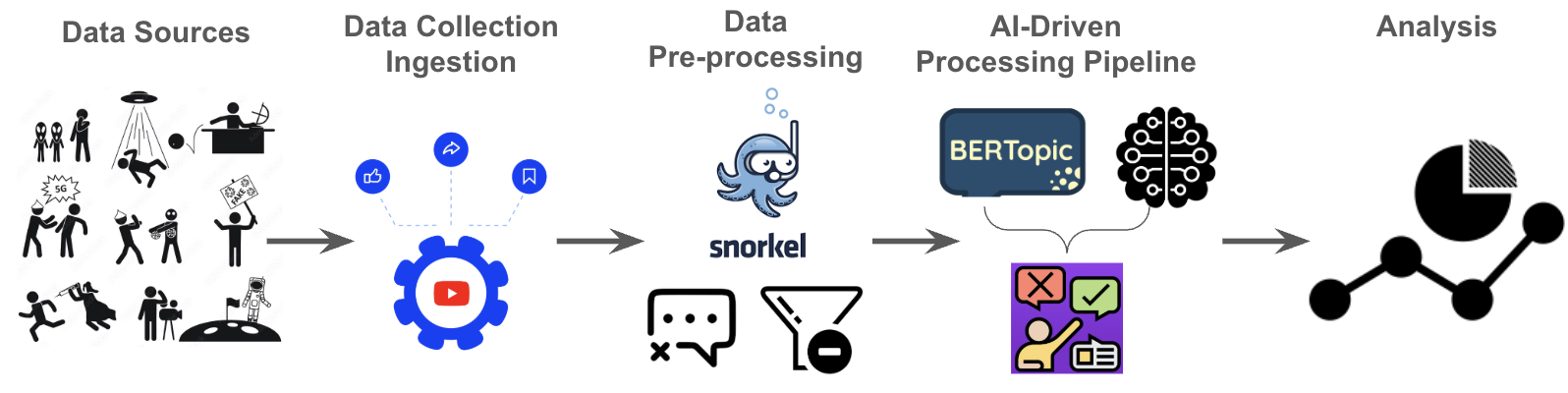}
\caption{ \small 
Overview of the proposed scalable AI-driven service architecture. 
The system consists of five layers: (1) Data Sources Curation, (2) Data Ingestion (YouTube API~\cite{youtube_api}), 
(3) Processing Pipeline (filtering, topic modelling, sentiment and stance analysis), 
(4) Analytics Layer (engagement metrics and behavioural signals), and 
(5) Service Interface for real-time querying and monitoring.
}
\label{fig.stance_pipeline}   
\vspace{-0.45cm}
\end{figure*}

The proposed framework consists of five main components: (\textit{i}) ground truth data source collection, (\textit{ii}) data ingestion, (\textit{iii}) preprocessing and filtering, (\textit{iv}) AI-driven processing pipeline, and (\textit{v}) analytics and service interface, as illustrated in Figure~\ref{fig.stance_pipeline}. The system is designed to operate on large-scale datasets and can be extended to support continuous data streams and real-time monitoring use cases.

\subsection{Ground Truth Data Sources}

\begin{table*}[h]
\centering
\small
\setlength{\tabcolsep}{4pt}
\begin{tabular}{l|p{2.1cm}|p{2.4cm}|p{2.7cm}|p{2.9cm}|p{1.8cm}}
\toprule
& \textbf{\# videos} 
& \textbf{\# videos with Comments} 
& \textbf{\# total comments in stats} 
& \textbf{\# total comments available to public} 
& \textbf{\# Unique Channels} \\
\hline
% \rowcolor{lightgray}
\multicolumn{6}{l}{\textbf{Dataset in \cite{Ballard2022}}}\\
\hline
Other Conspiracies & 66,956 & -- & -- & -- & --\\
QAnon & 7,281 & -- & -- & -- & --\\
\hline
% \rowcolor{lightgray}
\multicolumn{6}{l}{\textbf{Before June '23 policy update (round 1)}}\\
\hline
Other Conspiracies 
& 49,699 (74.2\%) 
& 47,063 
& 7,647,078 
& 6,654,277 
& 599 \\
QAnon 
& 2,210 (30.3\%) 
& 2,049 
& 641,576 
& 554,594 
& 50 \\
\hline
% \rowcolor{lightgray}
\multicolumn{6}{l}{\textbf{After June '23 policy update (round 2)}}\\
\hline
Other Conspiracies 
& 44,135 
& 41,231 
& 7,895,016 
& 7,340,029 
& 555 \\
QAnon 
& 2,394 
& 2,275 
& 787,329 
& 727,782 
& 45 \\
\bottomrule
\end{tabular}
\caption{\small Summary of the augmented dataset across different data collection rounds. The initial data sources are from \cite{Ballard2022}, containing \textit{only} video IDs and category labels.}
\label{tbl:datasetsummary}

\end{table*}
% \begin{table*}[h]
% \centering
% \begin{tabularx}{0.9\textwidth}{l|X|X|X|X|X}
% \toprule
%          &  \textbf{\# videos} & \textbf{\# videos with Comments} & \textbf{\# total comments in stats} & \textbf{\# total comments available to public} & \textbf{\# Unique Channels}\\
% \hline
% \rowcolor{lightgray}
% \multicolumn{6}{l}{\textbf{Dataset in \cite{Ballard2022}}}\\
% \hline
% Other Conspiracies & 66,956 & -- & -- & -- & --\\
% QAnon & 7,281 & -- & -- & -- & --\\
% \hline
% \rowcolor{lightgray}
% \multicolumn{6}{l}{\textbf{Before June '23 policy update (round 1)}}\\
% \hline
% Other Conspiracies & 49,699 (74.2\%) & 47,063 & 7,647,078 & 6,654,277 & 599 \\
% QAnon & 2,210 (30.3\%) & 2,049 & 641,576 & 554,594 & 50\\
% \hline
% \rowcolor{lightgray}
% \multicolumn{6}{l}{\textbf{After June '23 policy update (round 2)}}\\
% \hline
% Other Conspiracies & 44,135 & 41,231 & 7,895,016 & 7,340,029 & 555\\
% QAnon & 2,394 & 2,275 & 787,329 & 727,782 & 45\\
% \bottomrule
% \end{tabularx}
% \caption{\small Summary of the augmented dataset across different data collection rounds. The initial data sources are from \cite{Ballard2022}, containing \textit{only} video IDs and category labels.}
% \label{tbl:datasetsummary}
% % \vspace{-0.35cm}
% \end{table*}

Constructing large-scale datasets for conspiracy analysis is inherently challenging due to the lack of labelled data and the dynamic nature of online content. To address this, we build on the annotated dataset from \cite{Ballard2022} by adding additional metadata, comments, and engagement statistics.

The dataset consists of 66,956 ``Other Conspiracies'' videos and 7,281 QAnon videos published between 2007 and 2021. Using the YouTube Data API~\cite{youtube_api}, we collect publicly available comments, replies, likes, views, and metadata such as publication timestamps and channel identifiers. Additionally, we extract video transcripts using the YouTube-transcript-API~\cite{youtube-transcript-api}.

To capture platform dynamics, we perform two rounds of data collection. The first round (April 2023) provides a baseline snapshot, while the second round captures the impact of policy changes related to election misinformation. Table~\ref{tbl:datasetsummary} summarises the dataset characteristics.

\begin{table}[t]
\centering
\footnotesize
\caption{Mundane keywords for baseline dataset collection.}
\label{tab:mundane_keywords}
\begin{tabular}{lll}
\hline
song & school & book \\
cooking & surfing & finance \\
news & education & sports \\
health & comedy & 4WD \\
camping & kayaking & travel \\
people & plants & meal prep \\
day in my life & dance & movie \\
festival & ceremony & Christmas \\
thanksgiving & TED & hair care \\
beauty products & & \\
\hline
\end{tabular}
\end{table}
\noindent\textbf{Baseline Dataset:}

We accumulate a baseline dataset by querying the YouTube platform (through YouTube API \cite{youtube_api}) with a random set of 30 different ``mundane'' keywords (see Table~\ref{tab:mundane_keywords}) and scraping the top 100 most related videos. In this process, we set the ``video publication date'' parameter in a window between 2019 and 2021 (to align with the dates of publication of 94\% of the conspiracy videos in our initial ``Conspiracy videos'' dataset). This baseline dataset contains 379,722 user comments for 2,944 videos from 2,158 unique channels. We believe that the collected dataset of 3,000 ``random'' videos is representative of mainstream, non-conspiracy content. Although we do not manually verify the absence of conspiracy content, this dataset serves as a reasonable proxy for typical platform activity.

\begin{figure}[h]
\centering
\includegraphics[width=0.9\linewidth]{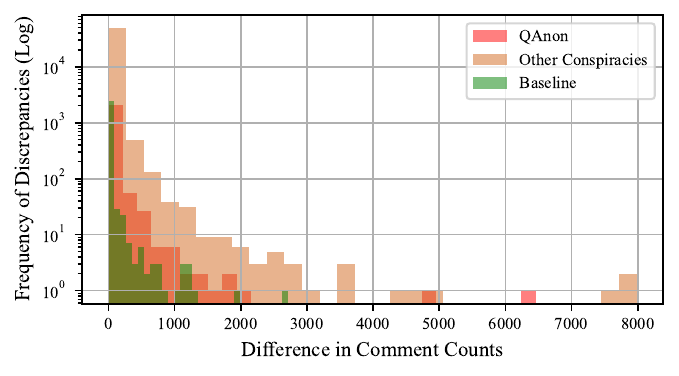}
\vspace{-0.25cm}
\caption{\small Differences between reported and publicly available comment counts across datasets (log scale).}
\label{fig.hist_descrepency}    
\vspace{-0.35cm}
\end{figure}

\noindent\textbf{Data Inconsistencies:}
We observe discrepancies between reported and publicly available comment counts across all datasets (Figure~\ref{fig.hist_descrepency}). Approximately 30.1\% of videos from the original dataset are no longer accessible, likely due to removal or moderation. Additionally, missing comment rates range from 9.8\% to 13.6\%, reflecting platform moderation and data availability constraints.

\subsection{Data Ingestion}

We implement a data ingestion module that collects content and engagement signals via the YouTube Data API~\cite{youtube_api}. The system processes both top-level comments and reply threads to capture complete interaction structures.

To support large-scale data collection, the ingestion process is implemented as a batched pipeline that handles API rate limits and ensures fault-tolerant data retrieval. The collected data is normalised into a unified schema, linking video identifiers, user identifiers, timestamps, and interaction metadata. This structured representation facilitates efficient downstream processing and analytics.

To ensure accurate measurement of user engagement, we exclude comments generated by content creators (i.e., channel owners). This step is critical to avoid bias introduced by self-engagement or promotional interactions. After filtering creator-generated content, the resulting dataset consists of approximately 6,426,665 comments for Other Conspiracies and 543,028 comments for QAnon videos (see Table~\ref{tbl:datasetsummary}), representing large-scale audience-driven engagement.

\subsection{Processing and Filtering}

Given the noisy and unstructured nature of user-generated content, we design a preprocessing and filtering module to improve data quality before downstream analysis. Social media comments frequently contain spam, promotional content, hyperlinks, emojis, and low-information expressions (e.g., ``lol'', ``haha''), which can negatively impact NLP-based inference tasks.

We formulate irrelevant comment detection as a binary classification problem and adopt a weakly supervised learning approach using the Snorkel framework~\cite{ratner2017snorkel}. We define \textit{irrelevant comments} as those matching one or more of the following patterns: (\textit{i}) single characters or emojis without meaningful semantic content, (\textit{ii}) standalone URLs or promotional phrases (e.g., ``subscribe'', ``donate''), (\textit{iii}) Ambiguous short expressions (e.g., ``lol'', ``haha''), and (\textit{iv}) spam or self-promotion requests (e.g., ``support my channel'')

Conversely, short but contextually meaningful comments (e.g., ``agree'', ``great point'', ``thank you'') are retained, as they often convey stance or sentiment.

To operationalise this, we perform corpus-level analysis using ``text locate'' function in the R \texttt{Corpus} package~\cite{rcorpus} to identify frequent linguistic patterns such as bigrams, repeated tokens, and link-heavy content. Based on these observations, we design a set of labelling functions that capture heuristic signals including URL presence, repetition, token sparsity, and lexical patterns. 
These labelling functions are combined using a generative model to produce probabilistic labels, which are then used to train a discriminative classifier. This weak supervision approach eliminates the need for large-scale manual annotation while maintaining strong classification performance. 
The filtering model achieves an F1-score of 0.95 on a manually annotated evaluation dataset, demonstrating its effectiveness in removing irrelevant and low-information comments. This preprocessing step ensures that subsequent modules operate on high-quality, context-rich textual inputs.

\subsection{AI-Driven Processing Pipeline}

The core of our system is a modular AI-driven processing pipeline that performs multi-stage analysis, including topic modelling, sentiment and emotion classification, and stance detection. The pipeline is designed to operate on large-scale datasets and can be integrated into service-oriented architectures for continuous and real-time analytics.

\textbf{Topic Modelling:}  
To capture the contextual semantics of videos, we apply BERTopic~\cite{grootendorst2022bertopic} to video transcripts. Since transcripts are automatically generated and often lack proper punctuation and sentence boundaries, we first perform text normalisation. This includes sentence reconstruction using a punctuation restoration model (Python Punctuator) and removal of frequent non-content phrases (e.g., disclaimers, channel introductions), identified through corpus analysis using the R \texttt{Corpus} package~\cite{rcorpus}. We manually verify that removing such phrases does not alter the semantic meaning of transcripts.

Unlike traditional topic modelling approaches based on word-frequency matrices, our objective is to extract human-interpretable topics that preserve semantic relationships within the data. Therefore, we adopt a transformer-based embedding approach. Specifically, we generate document embeddings using the pre-trained sentence transformer \textit{all-mpnet-base-v2}, which is well-suited for capturing semantic similarity and paraphrase relationships in long-form text.

The embedding space is then reduced using UMAP~\cite{mcinnes2018umap}, followed by density-based clustering using HDBSCAN~\cite{mcinnes2017hdbscan}. This enables the identification of fine-grained and semantically coherent clusters corresponding to distinct conspiracy topics. We fine-tune the parameters of UMAP and HDBSCAN using clustering quality metrics, including the Silhouette Score and Davies-Bouldin Index.

\begin{figure*}
\centering
\includegraphics[width=0.7\textwidth]{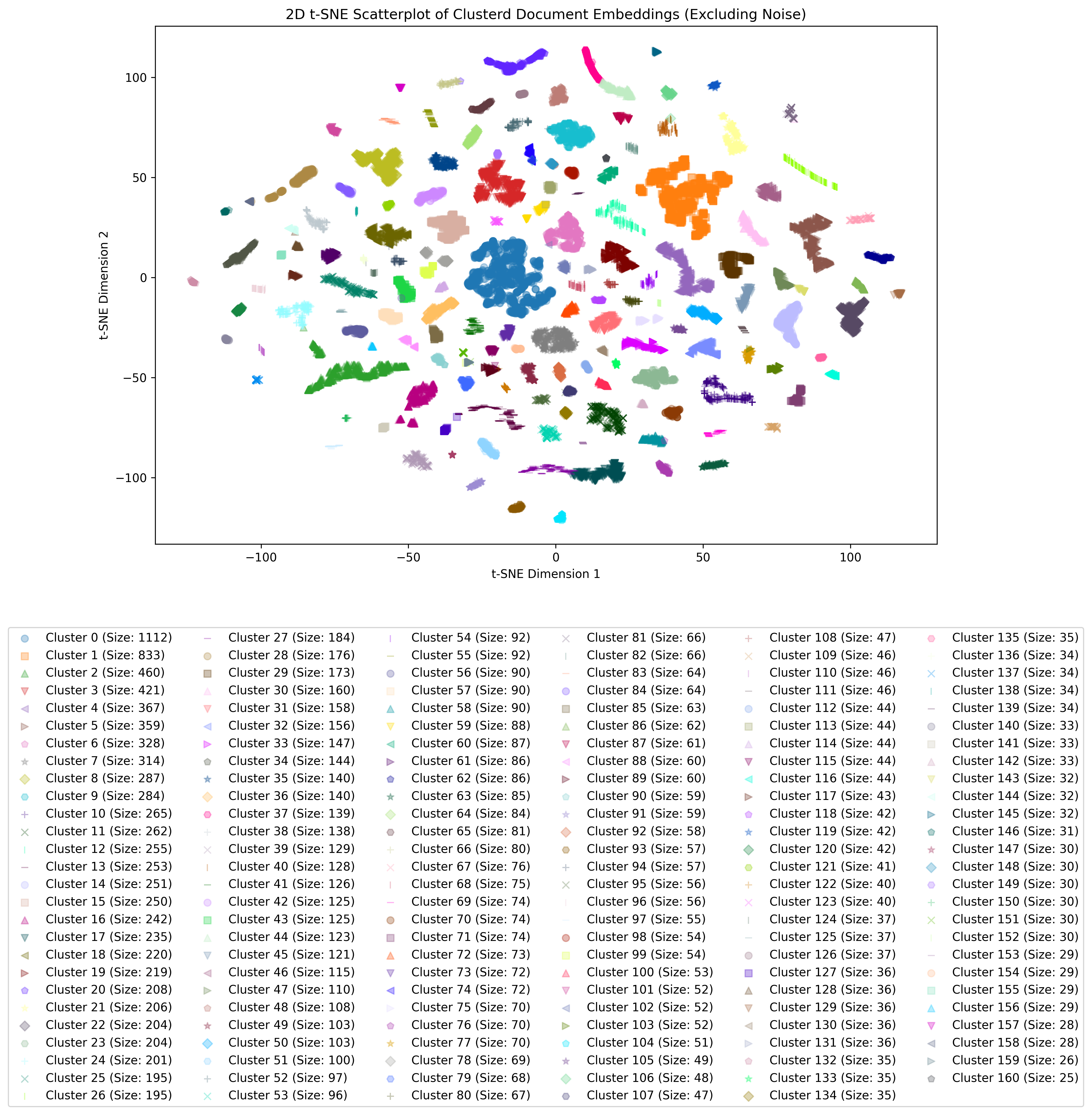}
\vspace{-0.15cm}
\caption{\small Clustered document embeddings showing coherent topic clusters.}
\label{fig:2D_t_SNE}
% \vspace{-0.5cm}
\end{figure*}

To further validate the clustering quality, we visualise the reduced embeddings using t-SNE (Figure~\ref{fig:2D_t_SNE}), which shows well-separated clusters. Quantitatively, the clustering achieves a Silhouette Score of 0.5887 and a Davies-Bouldin Index of 0.3976, indicating strong cohesion and separation.

After clustering, BERTopic applies a vectorizer model~\cite{scikit-learn} to extract representative keywords for each topic. We extend the standard NLTK stop-word list~\cite{loper2002nltk} with corpus-specific stop-words identified during preprocessing. We use an n-gram range of 2–3 to extract meaningful phrases that better represent topic semantics.

We evaluate topic interpretability using the Gensim coherence model~\cite{rehurek_lrec} and assess inter-topic similarity using cosine similarity. Tables~\ref{tbl:longother} and~\ref{tbl:longqanon} present the most coherent topics (coherence $> 0.6$), which correspond to real-world themes such as COVID-19, geopolitics, and financial systems.

\begin{table*}[ht]
\centering
\small
\caption{\small Topics in Other Conspiracies Dataset (Coherence $> 0.6$).}
\label{tbl:longother}
\vspace{-0.25cm}
\begin{tabular}{p{1cm} p{9cm} p{1.5cm} p{1.5cm} p{1.5cm}}
\hline
Topic & Keywords (Top Terms) & Coh. & Videos & Comments \\
\hline
0 & corona virus; bio weapon; death rate; public health; mortality rate & 0.75 & 1,106 & 195,400 \\
1 & great pyramid; ancient egypt; lost civilization; giza plateau & 0.85 & 830 & 149,112 \\
2 & super bowl; solar eclipse; kobe bryant; lebron james & 0.77 & 458 & 80,593 \\
3 & higher consciousness; quantum field; human collective; energy & 0.81 & 419 & 97,261 \\
4 & ufo sightings; space force; flying saucer; space station & 0.78 & 366 & 69,414 \\
6 & area 51; crop circles; ufo phenomenon; bob lazar & 0.78 & 327 & 53,352 \\
8 & spiritual themes; healing; soul mate; belief systems & 0.80 & 286 & 57,133 \\
9 & ascension; fifth dimension; divine plan; energy field & 0.81 & 283 & 40,631 \\
12 & aerial phenomena; defense; ufo disclosure; national security & 0.76 & 255 & 47,133 \\
16 & joe biden; elections; voter fraud; democratic party & 0.63 & 242 & 33,375 \\
\hline
\end{tabular}
% \vspace{-0.35cm}
\end{table*}

\begin{table*}[h]
\centering
\small
\caption{Topics in QAnon Dataset (Coherence $> 0.6$).}
\label{tbl:longqanon}
\vspace{-0.25cm}
\begin{tabular}{p{1cm} p{9cm} p{1.5cm} p{1.5cm} p{1.5cm}}
\hline
Topic & Keywords (Top Terms) & Coh. & Videos & Comments \\
\hline
0 & justice department; hillary clinton; voter fraud; white house & 0.85 & 173 & 83,797 \\
1 & roman empire; bible; holy spirit; kingdom of god & 0.82 & 131 & 43,829 \\
4 & president trump; oval office; national security & 0.75 & 78 & 12,249 \\
5 & deep state; fake media; surveillance; general flynn & 0.85 & 71 & 18,956 \\
6 & banking system; federal reserve; currency reset & 0.92 & 61 & 5,154 \\
9 & oil and gas; finance; hedge funds; markets & 0.89 & 34 & 3,376 \\
12 & healthcare; pharmaceuticals; immune system; disease & 0.77 & 24 & 3,737 \\
13 & deep state; obama administration; political narratives & 0.68 & 21 & 3,743 \\
\hline
\end{tabular}
\vspace{-0.35cm}
\end{table*}

\textbf{Sentiment and Emotion Analysis:}  
We use a pre-trained RoBERTa-based model~\cite{SemEval2018Task1} to classify user comments into sentiment and emotion categories. The model is trained on social media corpora, enabling it to capture nuanced linguistic phenomena such as sarcasm, informal language, and emotional intensity. These signals provide complementary behavioural insights that enhance the interpretation of user engagement patterns.

\noindent \textbf{Stance Detection on Comments:}

Stance detection is inherently a context-dependent task, requiring an understanding of the topic or narrative to which a user is responding in order to accurately infer their opinion. In large-scale social media environments, this challenge is further compounded by the presence of noisy, ambiguous, and low-information content. In particular, spam-like, irrelevant, or meaningless comments are prevalent and can negatively impact stance inference. To address this, we first preprocess the comment data by removing irrelevant and low-information content using the filtering module described earlier. 

\begin{figure*}[!th]
\centering
\includegraphics[width=0.9\textwidth]{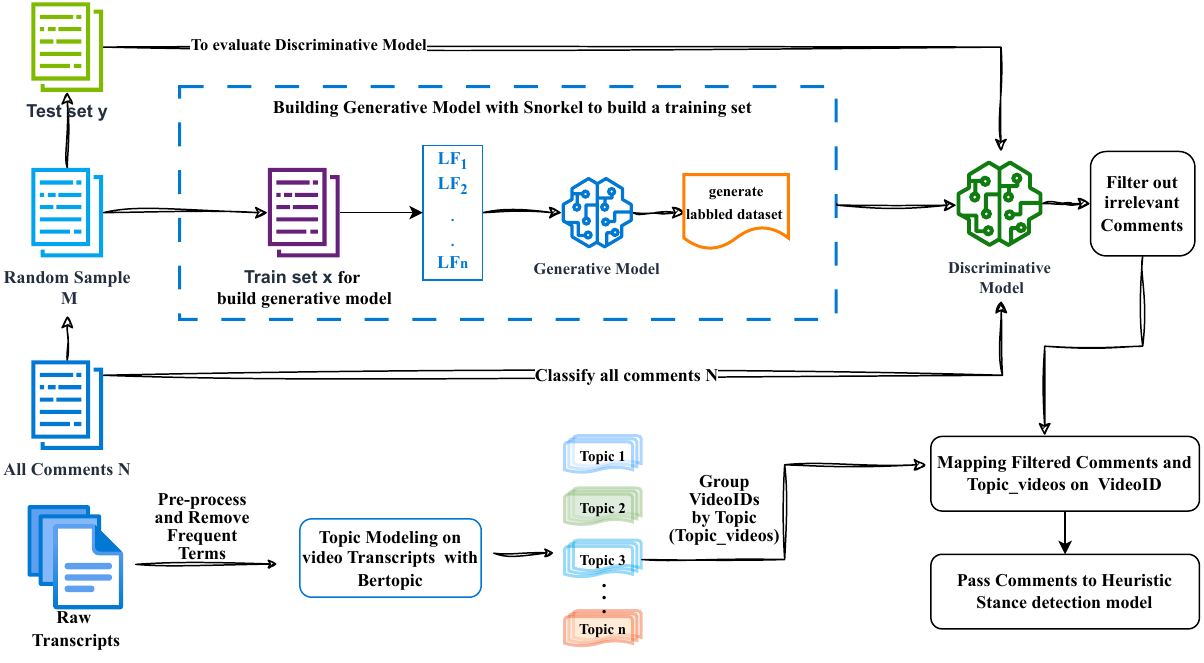}
\vspace{-0.25cm}
\caption{\small Our data preprocessing pipeline for stance detection.}
\label{fig.stance_pipeline1}   
% \vspace{-0.40cm}
\end{figure*}

To establish contextual grounding for stance detection, we then identify fine-grained conspiracy topics through topic modelling on video transcripts. These topics provide the semantic context required to interpret user comments. Figure~\ref{fig.stance_pipeline1} illustrates the overall pipeline, including comment filtering and context identification for stance detection.

The field of stance detection in social media continues to evolve, with most state-of-the-art (SOTA) models achieving accuracies between 60\% and 75\% on benchmark datasets. However, applying these approaches to large-scale, real-world data introduces several challenges, including (\textit{i}) the absence of labelled datasets, (\textit{ii}) the ambiguity and informality of user-generated text, and (\textit{iii}) the dynamic evolution of user opinions within discussion threads.

While prior work has largely relied on supervised learning approaches~\cite{rochert2022caught}, such methods are not directly applicable in this context due to the lack of annotated datasets and the high cost and ethical concerns associated with manual labelling of conspiracy-related content. This motivates the need for scalable, weakly supervised alternatives.

Building on the semi-supervised approach proposed in~\cite{reveilhac2023replicable}, we design a hybrid heuristic stance detection module that integrates semantic similarity, knowledge-based reasoning, and rule-based linguistic features. This approach extends prior work by incorporating contextual signals derived from topic modelling and domain-specific knowledge representations.

For top-level comments, stance detection is performed in two stages. First, the system checks for explicit expressions of agreement or disagreement (e.g., direct affirmation or rejection). If no explicit signal is detected, the comment is processed using an ensemble inference mechanism that combines multiple signals:

\begin{itemize}[topsep=0pt]
    \item \textbf{Semantic Similarity:} We compute the similarity between the comment and the topic representation derived from BERTopic to assess alignment with the underlying narrative.
    
    \item \textbf{Knowledge Base Entailment:} We construct lightweight domain-specific knowledge bases for prominent topics and evaluate whether the comment supports or contradicts key claims associated with the topic.
    
    \item \textbf{Rule-based Linguistic Features:} We incorporate heuristic signals such as negation patterns, antonyms near topic keywords, excessive punctuation, and uppercase emphasis, which often indicate strong stance or opposition.
\end{itemize}

These signals are combined to infer the final stance label (i.e., \textit{favour}, \textit{against}, or \textit{neutral}).

For reply comments, we extend stance detection by modelling conversational dependencies. Comments are grouped by their parent (top-level) comment and processed in temporal order. Stance is inferred by analysing the relationship between replies and their corresponding parent comments. Specifically, if a reply contradicts a parent comment expressing a favourable stance, it is classified as opposing, and vice versa. Additionally, when replies reference specific users within the thread, we treat the referenced comment as the contextual anchor for stance inference. This enables contextual stance propagation and captures evolving user opinions within discussions.

\begin{table}[!h] 
% \vspace{-0.15cm}
\centering 
\begin{tabular}{lcccc} \toprule & \textbf{Precision} & \textbf{Recall} & \textbf{F1-Score} & \textbf{Support} \\ \midrule Against & 0.12 & 0.33 & 0.18 & 12 \\ Favour & 0.95 & 0.84 & 0.89 & 180 \\ \midrule Accuracy & & & 0.81 & 192 \\ Macro Avg & 0.54 & 0.59 & 0.53 & 192 \\ Weighted Avg & 0.90 & 0.81 & 0.85 & 192 \\ \bottomrule \end{tabular} 
\caption{\small Classification Report for the Evaluation Dataset - Other Conspiracies.} 
\label{tab:classification_report_stance_other} 
\vspace{-0.25cm}
\end{table}

\begin{table}[!th] 
% \vspace{-0.25cm}
\centering 
\begin{tabular}{lcccc} \toprule \textbf{Label} & \textbf{Precision} & \textbf{Recall} & \textbf{F1-Score} & \textbf{Support} \\ \midrule Against & 0.28 & 0.53 & 0.37 & 17 \\ Favour & 0.89 & 0.74 & 0.81 & 92 \\ \midrule Accuracy & & & 0.71 & 109 \\ Macro avg & 0.39 & 0.76 & 0.39 & 109 \\ Weighted avg & 0.80 & 0.71 & 0.74 & 109 \\ \bottomrule \end{tabular} 
\caption{\small Classification Report for the Evaluation Dataset - QA-non Conspiracies.} \label{tab:classification_report_stance_qanon} 
\vspace{-0.15cm}
\end{table}

Tables~\ref{tab:classification_report_stance_other} and~\ref{tab:classification_report_stance_qanon} present the evaluation results of the stance detection module. The model achieves strong performance on favourable stance detection, with an overall accuracy of 0.81 for Other Conspiracies and 0.71 for QAnon datasets. The results further indicate that the majority of users express favourable stances toward conspiracy narratives, with 68\% in the Other Conspiracies dataset and up to 87.4\% in the QAnon dataset, while only a small proportion of comments express opposing views.

Overall, the proposed approach provides a scalable and practical solution for stance detection in large-scale social media datasets, prioritising deployability and robustness in the absence of labelled data.

{\bf Summary and Takeaway.} The processing and filtering pipeline transforms raw, noisy social media data into structured, high-quality inputs suitable for large-scale analytics. By integrating weak supervision, semantic modelling, and hybrid inference techniques, the system achieves both scalability and robustness. 
This design enables deployment in service-oriented environments, supporting real-time monitoring, behavioural analysis, and decision-making for applications such as content moderation, misinformation detection, and platform governance.

\section{Analysis and Results}
\label{sec:analysi}

In this section, we present the evaluation of our framework by addressing the three research questions defined in Section~\ref{sec:intro}. Before analysis, we remove comments generated by content creators to focus exclusively on audience engagement. This results in 6,426,665 comments for Other Conspiracies and 543,028 comments for QAnon videos.

\subsection{$\mathbf{RQ1}$: Engagement Analysis}
\label{subsec:RQ1}

% \textbf{RQ1: How do users engage with conspiracy content compared to mainstream content in terms of interaction patterns such as comments and likes?}

To answer $\mathbf{RQ1}$, we analyse user engagement patterns across conspiracy and baseline datasets using statistical and distribution-based methods.

\noindent \textbf{(\textit{i}) Distribution of Engagement:}
We first examine whether engagement distributions differ across conspiracy categories. To this end, we apply the Mann--Whitney U test~\cite{mann1947test}, a non-parametric statistical test suitable for comparing distributions without assuming normality. The null hypothesis assumes no difference between the distributions of comments received by videos in the two datasets.

The test yields an extremely low \textit{p}-value of $1.03 \times 10^{-10}$, leading to a rejection of the null hypothesis. This result provides strong statistical evidence that the comment distributions for Other Conspiracies and QAnon videos are not drawn from the same underlying distribution. In practical terms, this indicates that different categories of conspiracy content attract fundamentally different user engagement patterns. As a result, we analyse these datasets separately in subsequent evaluations.

\begin{figure*}
\centering
\begin{minipage}[t]{0.31\textwidth}
\centering
\includegraphics[width=\linewidth]{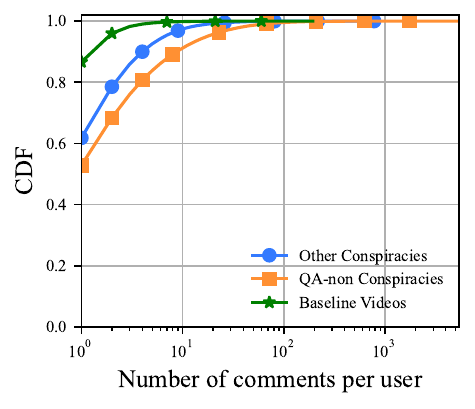}
\vspace{-0.25cm}
\caption{\small Comments distribution.}
\label{fig:cdf_comments_per_user}
\end{minipage}
% \hspace{3pt}
\begin{minipage}[t]{0.31\textwidth}
\centering
\includegraphics[width=\linewidth]{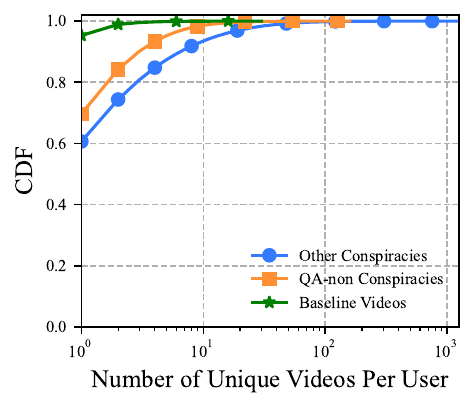}
\vspace{-0.25cm}
\caption{\small Unique videos a user has engaged.}
\label{fig:videos_per_user}  
\end{minipage}
% \hspace{3pt}
\begin{minipage}[t]{0.31\textwidth}
\centering
\includegraphics[width=\linewidth]{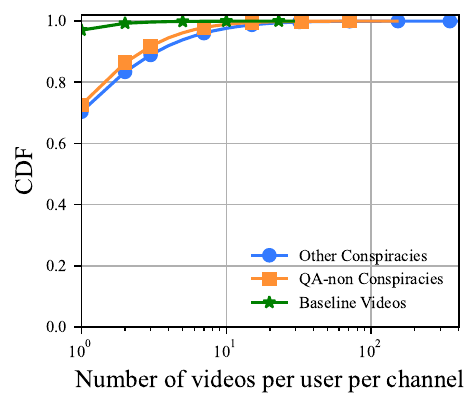}
\vspace{-0.25cm}
\caption{\small Unique videos per user by channel.}
\label{fig:videos_per_user_per_channel}  
\end{minipage}
\vspace{-0.35cm}
\end{figure*}

\noindent \textbf{(\textit{ii}) User Activity Patterns:}
To characterise individual user engagement, we analyse the distribution of comments per user across datasets. Figure~\ref{fig:cdf_comments_per_user} shows that all datasets exhibit highly right-skewed distributions, indicating that the majority of users contribute only a small number of comments, while a minority of users are highly active.

However, this skewness is significantly more pronounced in conspiracy datasets. For example, the maximum number of comments contributed by a single user reaches 5,437 in the Other Conspiracies dataset and 783 in QAnon, compared to only 194 in the baseline dataset. This suggests the presence of a subset of highly active users who disproportionately contribute to engagement in conspiracy content.

This observation is further supported by quartile analysis. The third quartile ($\mathbf{Q3}$) of comment counts is 2 and 3 for Other Conspiracies and QAnon, respectively, whereas it is only 1 for the baseline dataset. This indicates that a larger fraction of users in conspiracy datasets exhibit elevated engagement levels.

We also analyse repeated engagement at the video level by examining the number of comments posted by individual users on the same video. We observe extreme cases where a single user contributes up to 320 comments on a single video in the Other Conspiracies dataset and 172 comments in QAnon. Such repeated interactions suggest persistent engagement behaviour, potentially reflecting strong user interest, reinforcement dynamics, or coordinated activity.

\noindent \textbf{(\textit{iii}) Cross-Video Engagement:}
We next investigate whether users engage with multiple videos and whether such engagement is concentrated within specific channels. As shown in Figure~\ref{fig:videos_per_user}, users in the baseline dataset tend to engage with a limited number of videos, with the cumulative distribution rapidly approaching 1.0. This indicates that most baseline users interact with only one or a few videos.

In contrast, users engaging with conspiracy content interact with a significantly larger number of unique videos. This suggests a broader exploration or sustained interest in conspiracy-related topics.

To further analyse whether this engagement is channel-specific or content-driven, we examine the number of unique videos per user within individual channels (Figure~\ref{fig:videos_per_user_per_channel}). The results show that, while a small subset of users engages heavily, the majority of users do not concentrate their activity within a single channel. Instead, they interact with content across multiple channels.

This pattern indicates that engagement with conspiracy content is primarily driven by thematic interest rather than channel loyalty. In other words, users are more likely to follow the narrative across different sources rather than remain confined to a single content producer.

\noindent \textbf{(\textit{iv}) Interaction Correlation:}
To further understand the relationship between different forms of user engagement, we analyse the pairwise correlations between views, comments, and likes across videos in each dataset. This allows us to assess whether different engagement signals co-occur and whether highly engaging content exhibits consistent interaction patterns.
\begin{figure*}[!t]
\vspace{-0.35cm}
\centering
\subfloat[Other Conspiracies.]{%
    \includegraphics[width=0.32\textwidth]{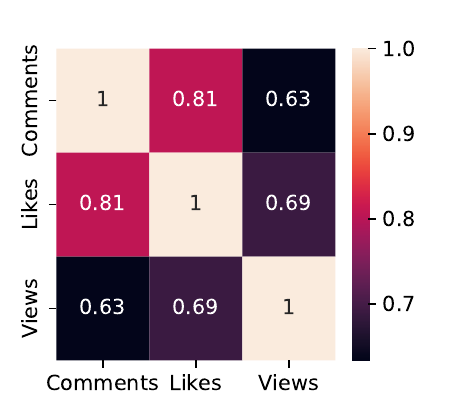}
    \label{fig:C_Correlation}}
\hfill
\subfloat[QAnon Conspiracy.]{%
    \includegraphics[width=0.32\textwidth]{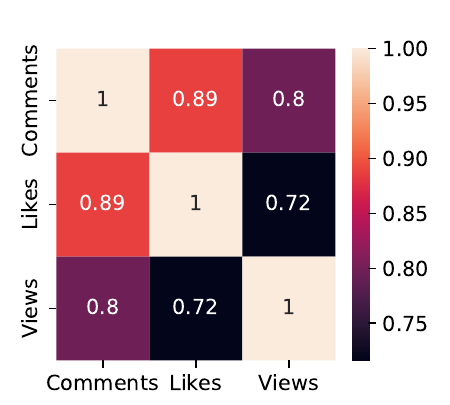}
    \label{fig:Q_Correlation}}
\hfill
\subfloat[Baseline Videos.]{%
    \includegraphics[width=0.32\textwidth]{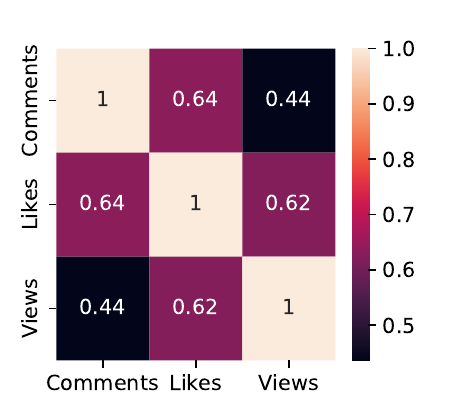}
    \label{fig:b_Correlation}}
\vspace{-0.25cm}
\caption{\small Pairwise Pearson correlation coefficient between users' number of comments, likes, and views in each dataset.}
\label{fig:stats_heatmaps}
% \vspace{-0.45cm}
\end{figure*}
Figure~\ref{fig:stats_heatmaps} presents the pairwise Pearson correlation coefficients between these engagement metrics. In the Other Conspiracies dataset, we observe a strong positive correlation between the number of comments and likes (Pearson coefficient = 0.81), indicating that videos attracting more discussion also tend to receive higher appreciation signals. Similarly, QAnon content exhibits comparable strong correlations across engagement metrics, suggesting cohesive interaction dynamics.

To validate the statistical significance of these relationships, we perform regression analysis, which yields a \textit{p}-value of $< 0.0001$. This indicates that the observed correlations are highly unlikely to be due to random chance. However, it is important to note that correlation does not imply causation. Due to the lack of user-level interaction traces, we cannot determine whether the same users who comment are also those who like the content.

In contrast, the baseline dataset exhibits noticeably weaker correlations between engagement metrics, as shown in Figure~\ref{fig:stats_heatmaps}. This suggests that user interactions with mainstream content are more heterogeneous and less tightly coupled across different engagement modes.

Overall, these findings indicate that conspiracy content tends to trigger more synchronised engagement behaviour, where multiple interaction signals (comments, likes, and views) reinforce each other. This cohesive engagement pattern may contribute to the amplification and visibility of such content on the platform.

\noindent \textbf{Takeaway ($\mathbf{RQ1}$):} Our findings suggest that conspiracy content exhibits stronger, more skewed, and more distributed engagement compared to mainstream content. A small subset of highly active users contributes disproportionately, and engagement is driven by content across multiple channels rather than isolated communities.

\subsection{$\mathbf{RQ2}$: Behavioural Signals}
\label{subsec:RQ2}

%\textbf{RQ2: Which sentiments, emotions, and stances influence active user engagement?}

To answer $\mathbf{RQ2}$, we analyse sentiment, emotion, and stance signals associated with highly active users (above $\mathbf{Q3}$ engagement threshold).

\begin{figure}[ht]
\centering
\vspace{-0.35cm}
\includegraphics[width=0.8\columnwidth]{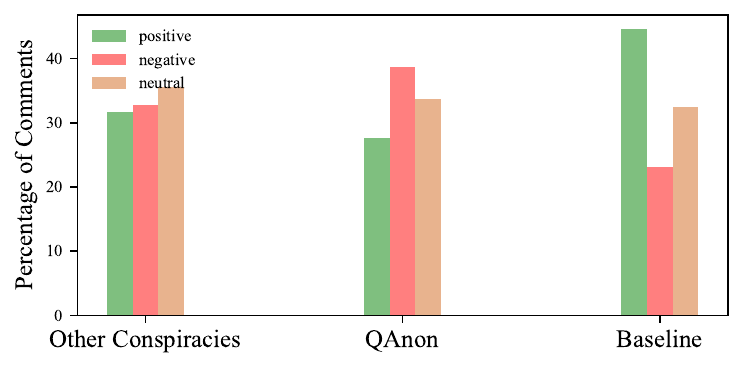}
\vspace{-0.25cm}
\caption{\small Proportion of the comments in each dataset by sentiments of most actively engaging users. }
% \Description{sentiment distribution of the dataset}
\label{fig.sentiments}   
% \vspace{-0.3cm}
\vspace{-0.35cm}
\end{figure}

\noindent \textbf{(\textit{i}) Sentiment Analysis:}
Figure~\ref{fig.sentiments} shows that active engagement in QAnon content is predominantly associated with negative sentiment, while Other Conspiracies show a mix of positive and negative sentiment. In contrast, baseline content is dominated by positive sentiment, indicating a fundamental difference in user behaviour across content types.

\begin{figure}[ht]
\centering
\includegraphics[width=0.8\columnwidth]{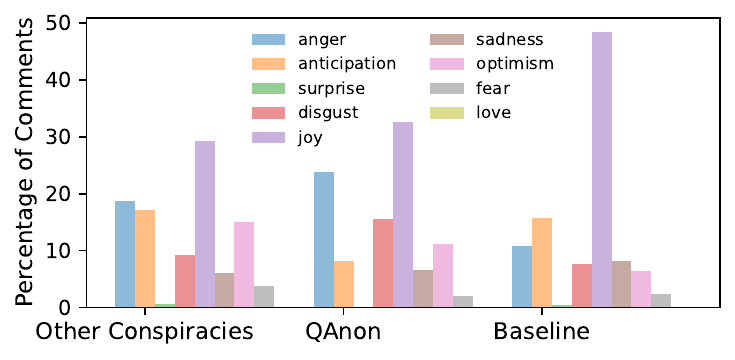}
\vspace{-0.25cm}
\caption{\small Sentiment distribution our dataset.}
% \Description{sentiment distribution of the dataset}
\label{fig.emotion}   
\vspace{-0.20cm}
\end{figure}

\noindent \textbf{(\textit{ii}) Emotion Analysis:}
Figure~\ref{fig.emotion} reveals that \textit{joy} is the most common emotion across all datasets, often reflecting appreciation toward content creators. However, conspiracy content shows a significantly higher proportion of \textit{anger}, indicating stronger emotional intensity and polarisation compared to mainstream content.

\noindent \textbf{(\textit{iii}) Stance Analysis:}
Our stance detection results (Tables~\ref{tab:classification_report_stance_other} and~\ref{tab:classification_report_stance_qanon}) show that the majority of users express favourable stances toward conspiracy narratives. Specifically, 68\% of comments in Other Conspiracies and 87.4\% in QAnon support the underlying narratives. This suggests that user engagement is not merely reactive but often aligned with the content's perspective.

\noindent \textbf{Takeaway ($\mathbf{RQ2}$):}
Active engagement with conspiracy content is characterised by stronger emotional signals, higher negative sentiment, and predominantly favourable stance toward the content. These behavioural signals indicate reinforcement dynamics that may contribute to the amplification of such narratives.

\subsection{$\mathbf{RQ3}$: Temporal Dynamics}
\label{subsec:RQ3}

% \textbf{RQ3: How does user engagement evolve over time following content publication?}

To answer $\mathbf{RQ3}$, we perform a longitudinal analysis of user engagement by leveraging comment timestamps. Specifically, we compute the time difference between each comment and the corresponding video publication time, enabling us to analyse how quickly users engage with content and how engagement evolves over time. This allows us to compare early-stage engagement dynamics as well as long-term temporal trends across conspiracy and baseline datasets.

\begin{figure}[h]
\centering
\includegraphics[width=0.75\columnwidth]{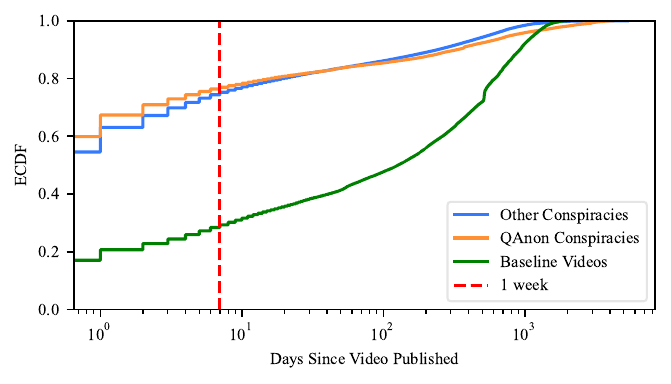}
\vspace{-0.25cm}
\caption{\small ECDF of user comments over time since the video published date.}
\label{fig.comments_over Time}   
% \vspace{-0.35cm}
\end{figure}

\begin{figure}[h]
\centering
\includegraphics[width=0.75\columnwidth]{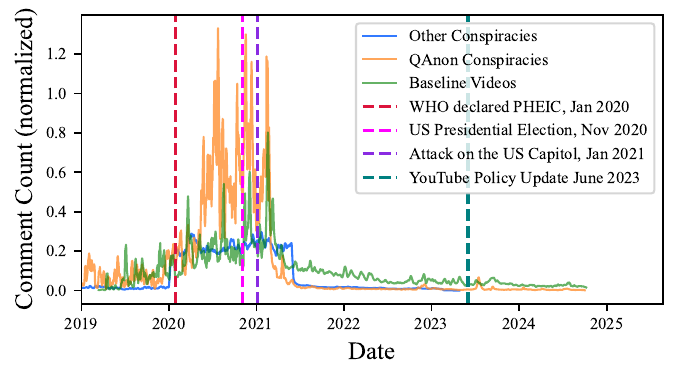}
\vspace{-0.25cm}
\caption{\small Time series analysis of comments received over time. Normalised comment count is obtained by dividing total comments by the number of videos, enabling comparison across datasets of different sizes.}
\label{fig.timeseries_comment_recieved}   
% \vspace{-0.35cm}
\end{figure}

\textbf{(\textit{i}) Early Engagement Dynamics:}  
Figure~\ref{fig.comments_over Time} presents the empirical cumulative distribution function (ECDF) of comment arrival times. The results show that conspiracy content attracts a substantial proportion of user engagement shortly after publication. Specifically, approximately 77\% of comments in the QAnon dataset and 75\% in the Other Conspiracies dataset occur within the first 7 days. In contrast, baseline content exhibits a slower accumulation of engagement over time.

This rapid early engagement suggests that conspiracy content benefits from strong initial visibility and user interest, potentially driven by recommendation systems, social sharing, or coordinated dissemination. Moreover, the steep rise in the ECDF curves indicates that engagement is highly concentrated in the early lifecycle of the content, highlighting a narrow window during which content gains traction.

\textbf{(\textit{ii}) Sustained and Long-Tail Engagement:}  
Beyond the initial surge, engagement patterns exhibit a long-tail distribution, where a smaller fraction of comments continues to accumulate over extended periods. However, compared to baseline content, the marginal growth in engagement for conspiracy videos declines more rapidly after the initial burst, suggesting that most interaction occurs during the early phase rather than being sustained over time.

\textbf{(\textit{iii}) Event-Driven Engagement Patterns:}  
Figure~\ref{fig.timeseries_comment_recieved} presents a time-series analysis of comment activity, normalised by the number of videos to ensure comparability across datasets. The results reveal distinct spikes in engagement, particularly in the QAnon dataset, which align with major real-world events (e.g., political developments or public crises). These spikes indicate that external events act as triggers that re-activate user engagement with existing conspiracy content.

This behaviour suggests that conspiracy narratives are not only consumed passively but are dynamically revisited and amplified in response to external stimuli, reinforcing their persistence within the platform.

\textbf{(\textit{iv}) Comparative Temporal Behaviour:}  
Compared to baseline content, conspiracy videos exhibit both faster initial engagement and stronger event-driven amplification. While baseline content shows relatively stable and gradual engagement patterns, conspiracy content demonstrates bursty and reactive dynamics, reflecting heightened sensitivity to external factors.

\textbf{Takeaway ($\mathbf{RQ3}$):}  
Conspiracy content is characterised by rapid early engagement, followed by limited long-term growth and strong responsiveness to real-world events. These temporal dynamics indicate that the early lifecycle of content is critical for amplification, and that external triggers can significantly influence engagement patterns. From a system perspective, these findings highlight the importance of early-stage detection and real-time monitoring services to mitigate the spread and resurgence of potentially harmful content.

\section{Limitations and Scope of Analysis}
\label{sec:limitations}

While our proposed framework enables scalable analysis of user engagement and behavioural signals, several limitations should be acknowledged.

Our analysis relies on publicly available data collected through the YouTube Data API. As a result, certain attributes—such as deleted comments, moderated content, user-level interaction histories, and private engagement signals—are not accessible. Additionally, we observe discrepancies between reported and publicly available comment counts, likely due to platform moderation or content removal. These factors may introduce bias and limit the completeness of the dataset.

Although our dataset spans a broad range of conspiracy-related content, it does not fully represent the entire YouTube ecosystem. Furthermore, our stance analysis focuses on the most prominent topics identified through topic modelling. As a result, the findings may not generalise to less prevalent or emerging conspiracy narratives.

Our stance detection module adopts a hybrid heuristic approach in the absence of large-scale labelled datasets. While this enables scalable inference, it requires domain knowledge to construct topic-specific knowledge bases and interpret contextual signals. Consequently, the effectiveness of the method depends on the availability of sufficient understanding of the underlying conspiracy narratives.

The evaluation of stance detection relies on manually labelled datasets, which introduces potential subjectivity, as annotators must interpret nuanced and often ambiguous content. Although we mitigate this by using multiple annotators and majority voting, the evaluation process may still reflect annotator bias and limited domain expertise.

Our temporal analysis captures engagement patterns based on observed timestamps but does not account for external recommendation algorithms or unseen exposure mechanisms that influence user behaviour. Therefore, while we identify strong correlations between events and engagement spikes, we do not claim causal relationships.

Due to these constraints, our stance analysis is limited to selected high-confidence topics within the datasets. While this ensures the reliability of the results, it restricts broader generalisation across all types of conspiracy content.

Despite these limitations, our framework provides a scalable and practical approach for analysing user engagement, behavioural signals, and stance dynamics in large-scale social media environments.

\section{Conclusion \& Future Work}
\label{sec.Conclusion and Future Work}

In this paper, we presented a scalable AI-driven framework for analysing user engagement, behavioural signals, and stance dynamics in conspiracy-related content on YouTube. By integrating data ingestion, weakly supervised filtering, topic modelling, sentiment analysis, and a hybrid stance detection module, our system enables large-scale, service-oriented analysis without requiring extensive labelled data.

Our evaluation provides three key insights. First, engagement with conspiracy content is highly skewed and driven by a subset of highly active users who interact across multiple videos and channels, indicating content-driven rather than channel-driven behaviour. Second, behavioural signals show that active engagement is associated with stronger emotional intensity, higher negative sentiment, and predominantly favourable stance toward conspiracy narratives, suggesting reinforcement dynamics within user communities. Third, temporal analysis reveals that engagement is heavily concentrated in the early lifecycle of content, with the majority of interactions occurring shortly after publication and further amplified by real-world events.

From a systems perspective, these findings highlight the need for scalable and real-time monitoring services that prioritise early-stage detection and dynamically respond to event-driven engagement surges. The proposed framework provides a practical foundation for building such services, enabling continuous analysis of large-scale user interactions and behavioural patterns. 
Despite these contributions, challenges remain due to the lack of large-scale labelled datasets and the complexity of social media text, which limit the applicability of fully supervised approaches and introduce ambiguity in stance interpretation.

In future work, we plan to extend our framework across multiple social media platforms to evaluate the generalisability of engagement and behavioural patterns. We also aim to explore transfer learning and active learning approaches to improve stance detection with minimal annotation effort. Additionally, integrating streaming data pipelines and deploying the system as a real-time service will enable proactive monitoring and intervention mechanisms for emerging misinformation and conspiracy narratives.

\balance
\bibliographystyle{IEEEtran}

\bibliography{references}

\end{document}